\documentclass[preprint]{elsarticle}
\usepackage{amsmath}
\usepackage{amsfonts}
\usepackage{amssymb}
\usepackage{amsthm}
\usepackage{graphicx}
\usepackage{hyperref}
\usepackage{color}
\usepackage{flafter}
\usepackage[normalem]{ulem}
\newtheorem{define}{Definition}

\newenvironment{blockquote}{%
	\par%
	\medskip
	\leftskip=2em\rightskip=2em%
	\noindent\ignorespaces}{%
	\par\medskip}

\newcommand{\Fig}[1]{Fig.~\ref{#1}}

\newcommand{\Eq}[1]{equation~(\ref{#1})}
\newcommand{\nn}{\nonumber\\}
\newcommand{\ie}{{\emph{i.e.~}}}

\renewcommand{\>}{\rangle}

\newcommand{\be}{\begin{eqnarray}}
\newcommand{\ee}{\end{eqnarray}}
\newcommand{\bpm}{\begin{pmatrix}}
\newcommand{\epm}{\end{pmatrix}}

\newcommand{\Tr}{{\rm Tr}}

\newcommand{\ra}{\rightarrow}

\renewcommand{\v}[1]{{\boldsymbol{#1}}}
\renewcommand{\a}{\alpha}
\renewcommand{\b}{\beta}

\newcommand{\e}{\epsilon}
\newcommand{\s}{\sigma}

\newcommand{\G}{\Gamma}

\begin{document}

\title{The ``non-regularizability'' of gapless free fermion Hamiltonian protected by on-site symmetries}

\author[ucb]{Yen-Ta Huang}
\ead{yenta.huang@berkeley.edu}
\author[ucb]{Lokman Tsui}
\ead{lokman@berkeley.edu}
\author[ucb,lbl]{Dung-Hai Lee\corref{cor1}}
\ead{dunghai@berkeley.edu}
\cortext[cor1]{Corresponding author}
\address[ucb]{Department of Physics, University of California, Berkeley, California 94720, USA}
\address[lbl]{Materials Sciences Division, Lawrence Berkeley National Laboratories, Berkeley, California 94720, USA}

\setlength{\parindent}{0pt}

\date{\today}

\begin{abstract}
The non-regularizability of free fermion field theories, which is the root of various quantum anomalies, plays a central role in particle physics and modern condensed matter physics.  In this paper, we generalize the Nielsen-Ninomiya theorem to all minimal nodal free fermion field theories protected by the time reversal, charge conservation, and charge conjugation  symmetries. We prove that these massless field theories cannot be regularized on a lattice.
\end{abstract}

\maketitle

\section{Introduction} \label{sec:intro}

The non-regularizability of massless free fermion field theories is the origin of various quantum anomalies. A famous example  is the Nielsen-Ninomiya\cite{Nielsen1981} theorem, namely,  Weyl nodes with net chirality cannot be realized by any charge-conserved lattice model in three dimensions. However,  Weyl nodes with net  chirality can appear on the boundary of a 4D charge-conservation-protected topological insulator (The free fermion topological classification of this 4D topological insulator is $\mathbb{Z}$.). Another example involves Dirac cones with net vorticity in 2D. Under charge conservation and time-reversal ($ T^2=-1 $) symmetries, Dirac cones with net vorticity cannot be realized by any lattice model.  However, they can appear on the boundary of a 3D topological insulator. (The free fermion topological classification of such topological insulator is $\mathbb{Z}_2$.)\\

According to the folklore, the low energy field theory describing the boundary of on-site symmetry protected topological states (SPTs) cannot be regularized on a lattice. In other words, they can not be realized as finite-range tight-binding models where the symmetry acts  on the degrees of freedom on each lattice site independently. The obstruction lies in the realization of symmetry -- in the boundary dimension the {\underline{on-site}} nature of the protection symmetry cannot be realized. This obstruction is relieved by ``UV completing'' the boundary degrees of freedom with the bulk degrees of freedom living in one extra spatial dimension. The bulk degrees of freedom are gapped and respect an on-site symmetry. The bulk state is called an SPT.  When the boundry is one-dimensional, Ref.\cite{Ryu2012,Sule2013} argued that  the non-regularizability is manifested by the fact that the boundary theory is not modular invariant  after orbifolding with respect to the  protection symmetry. \\

If the protection symmetry is not on-site, regularization is certainly possible. A famous example is the tight-binding model of graphene. There, the two Dirac nodes are protected by the translation, charge conservation, time reversal, and inversion symmetries. Here the inversion symmetry is not on-site. In the rest of the paper we shall assume translation invariance and the term ``symmetry'' always refers to other on-site symmetry.\\

The non-regularizability discussed above lies at the heart of the physics of SPTs. It is well-known that  SPTs are defined by their symmetry-protected gapless boundaries. In the following, we argue that if it were possible to realize these boundaries on a lattice, the gapless boundary modes will not be protected.\\

\begin{figure}[h]
\begin{center}
\includegraphics[scale=0.35]{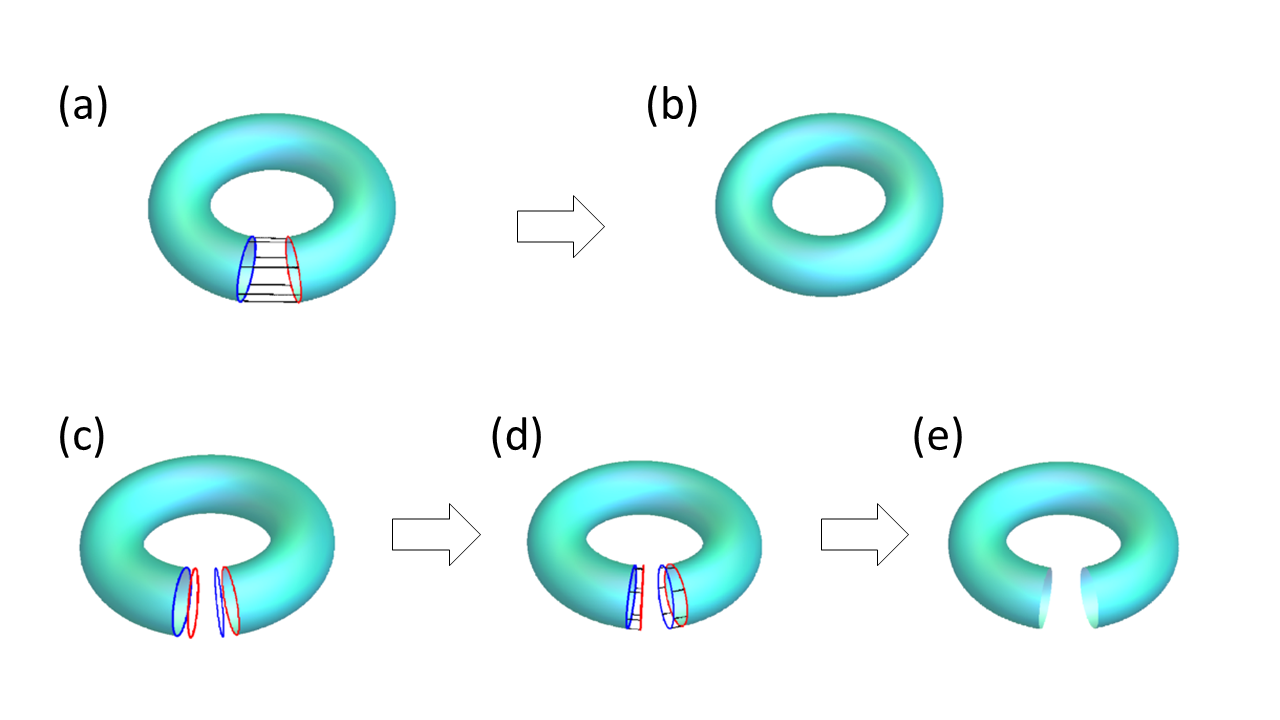}
\caption{An illustration of the fact that the regularizability of the boundary Hamiltonian of an SPT implies the gapless modes are not  protected. (a,b) By turning on symmetry-respecting interactions  mimicking those in the bulk (the black lines) it is possible to gap out the gapless modes (red and blue circles). (c) The regularizability of the boundary SPT Hamiltonian implies it is possible to fabricate the gapless boundaries. (d,e) The fabricated boundaries can be brought to interact with the original boundaries and gap each other out.    }
\label{fig1}
\end{center}
\end{figure}

For example, in \Fig{fig1} we consider a 2D SPT having two edges. It is always possible to reconnect these edges with symmetry-respecting interactions, i.e., seal off the boundary (\Fig{fig1}(a)). After the reconnection the gapless modes are  removed (\Fig{fig1}(b)). If it was possible to regularize the gapless boundaries on 1D lattices, one would have been able to fabricate the gapless boundaries as 1D systems (\Fig{fig1}(c)). These fabricated edges can be brought around to interact with the original boundaries (via symmetry-respecting interactions) (\Fig{fig1}(d)). As a result, the gapless edges can be removed (\Fig{fig1}(e)), which proves that the original gapless edges are not symmetry-protected.\\

The purpose of this paper is to prove the following folklore, namely:
\begin{blockquote}
\emph{Any symmetry-protected minimal nodal free-fermion field theory cannot be regularized on a lattice.}
\end{blockquote}
Here, ``nodal free-fermion field theory'' is a continuum field theory which has a gapless spectrum with a linear-dispersing gap node, characterized by a Clifford algebra, at a single time-reversal invariant  momentum. Without loss of generality, we shall assume such momentum to be $\v k = 0$. ``Minimal'' refers to the fact that the fermion field in the theory has the smallest number of components necessary to represent the symmetry transformations and the Clifford algebra. ``Symmetry-protection" means there is no symmetry-allowed mass term. In this paper, we restrict ourselves to the charge conservation, time-reversal, and charge conjugation symmetries. ``Lattice regularization" is the procedure which converts the continuum field theory to a finite-range tight-binding model while preserving all symmetries.\\

The outline of the paper is as follows. We achieve the proof by  ``reductio ad absurdum''. In section \ref{lat}, we assume the existence of a tight-binding Hamiltonian whose low energy limit is the field theory in question. Let the momentum space Hamiltonian of this tight-binding model be $h(\v k)$, we list the four constraints $h(\v k)$ must obey. In section \ref{form} we present the $h(\v k)$ which has the smallest matrix size and satisfies the constraints listed in section \ref{lat}. This is the momentum space Hamiltonian of the minimal models. In section \ref{symprot} we lay out the symmetry protection hypothesis.  In section \ref{symmetrise} and \ref{preserve}, we show that for each $h(\v k)$ obeying the constraints of sections \ref{lat} and \ref{symprot} there is an associated  ``spectral symmetrised" counterpart, $\tilde{h}(\v k)$. In section \ref{PH} we apply the Poincar\'e-Hopf Theorem to $\tilde{h}(\v k)$, and show that it imposes a stringent constraint on the form of $\tilde{h}(\v k)$ at a time reversal invariant point $\v k_0$ different from $\v k=0$. Section \ref{prf} adopts the strategy of reductio ad absurdum for the proof of non-regularizability. We complete the proof in two alternative ways. (a) When $\tilde{h}(\v k)$ satisfies a special condition we prove that if it obeys constraints 1-4 it must  violate the symmetry-protection hypothesis. (b) For other $\tilde{h}(\v k)$ we prove that if it satisfies the symmetry-protection hypothesis it must have the energy gap close at $\v k_0$ as well. This means it violates constraint 3 of section \ref{lat}. The proof (b) is achieved by a case-by-case study of all nodal Hamiltonians protected by the charge conservation, time reversal, and charge conjugation symmetries.  Because of the length of this proof, it is left to \ref{the proof} and \ref{odd_func}.

\section{The constraints on lattice-regularized nodal Hamiltonians }\label{lat}
\label{nodal}

In the following we  assume the existence of lattice-regularized minimal SPN Hamiltonian
\begin{align}
H=\sum_{\v k\in BZ}~\chi(-\v k)^T h(\v k) \chi(\v k), \label{hk}
\end{align}
and discuss the conditions it must satisfy.
Here  ``BZ'' stands for the Brillouin zone of a $d$-dimensional lattice.  $\chi(\v k)$ is a Fourier transformed  Majorana lattice field.  We work with Majorana rather than complex fermion field because it also covers charge non-conserving (Bogoliubov-de Gennes) free fermion Hamiltonians. 
There are 4 constraints we require 
$h(\v k)$ to satisfy:\\

\begin{enumerate}
\item The Majorana constraint: $h^{T}(-\v k) = -h(\v k)$.
\item $h(\v k)$ is an analytic function of $\v k$ in the Brillouin zone. 
\item There is an energy gap between the lower half and the upper half of the eigenvalues of $h(\v k)$. The energy gap between these two groups of eigenvalues exhibits a single node at $\v k=0$. Moreover
\be h(\v k)\ra \sum_{j=1}^d k_j \G_j ~{\rm as}~\v k\ra 0.\label{effh}\ee Here $\{\G_j\}$ are traceless symmetric matrices satisfying $\{\G_i,\G_j\}=2\delta_{ij}$.
\item $h(\v k)$ obeys the following symmetry requirement: $U_{\b}^{\dagger} h(\v k)U_{\b} = h(\v k)$ and  $A_{\a}^{\dagger} h(-\v k )^{*} A_{\a} = h(\v k)$. Here $A_\alpha,~\alpha=1,...,N_A$ and $U_\beta,~\b=1,...,N_U$ are $\v k$-independent orthogonal matrices representing the anti-unitary and unitary protection symmetries.
\end{enumerate}

Four comments are in order:
\begin{list} {$\circ$}{} 
\item We assume that the Hamiltonian has translation symmetry so that we can express it in momentum space. The more general case where the translation symmetry is absent is more diffucult, and is beyond the scope of this paper. The fact that the unitary and anti-unitary symmetry matrices do not depend on $\v k$ signifies that they are on-site symmetries.
\item In the presence of anti-unitary symmetry, $A_{\a}^{\dagger} h(-\v k )^{*} A_{\a} = h(\v k)$ implies $A_{\a}^{\dagger} h(\v k ) A_{\a} = -h(\v k)$ due to constraint 1.
As a result, the spectrum of $h(\v k)$ is symmetric about zero for each $\v k$. 
\item  In 1D we shall assume the dispersion of  $h( k)$ is non-chiral. This is because for chiral Hamiltonians the constraints of continuity, Brillouin zone periodicity, and the requirement that the energy band crosses the Fermi energy only at $k=0$ (which is the nodal condition  for chiral Hamiltonians) obviously contradict one another.
\item Conditions 3 and 4 impose a constraint on the minimal size of $h(\v k)$. We state, without proof, that the smallest such matrix for the charge conservation (unitary), time reversal (anti-unitary) and charge conjugation (unitary) symmetries has dimension $2^{n}\times 2^{n}$. Here $n$ depends on the spatial dimension and the symmetry group.
\end{list}

\section{The minimal model satisfying constraints 1-4 in section \ref{lat}}\label{form}

In this section and the rest of the paper we shall focus on minimal SPN models. For these models $h(\v k)$ is a $2^n\times 2^n$ Hermitian matrix. Any such $2^n\times 2^n$ $h(\v k)$ can be constructed from linear combinations of the tensor products of $n$ Pauli matrices.  
Among them $N_1=(2^{2n}+2^n)/2$ are real and symmetric and $N_2=(2^{2n}-2^n)/2$ are imaginary and anti-symmetric, i.e., 
\begin{align}
h(\v k) = \sum_{i=1}^{N_1} o_i(\v k) M^s_i + \sum_{j=1}^{N_2} e_j(\v k) M^a_j. \label{hmaj0}
\end{align}
 Due to the Majorana constraint $o_i(\v k)$ and $e_j(\v k)$ are odd and even functions of $\v k$, respectively. Under the action of unitary symmetries $\v k$ remains unchanged. But anti-unitary symmetries send $\v k$ to $-\v k$. As the result, the $\{M_i^s\}$  and $\{M_j^a\}$ that can appear in \Eq{hmaj0} must satisfy the following equations
\be
&&U_{\b}^{\dagger} M^s_i U_{\b} = M^s_i,~~A_{\a}^{\dagger} M^s_i A_{\a} =-M^s_i\nn
&&U_{\b}^{\dagger} M^a_i U_{\b} = M^a_i,~~A_{\a}^{\dagger} M^a_i A_{\a} =-M^a_i.\label{symm}\ee
Let the number of symmetric/anti-symmetric matrices satisfying \Eq{symm} be $n_s$ and $n_a$, respectively.  Thus
\begin{align}
h(\v k) = \sum_{i=1}^{n_s} o_i(\v k) M^s_i + \sum_{j=1}^{n_a} e_j(\v k) M^a_j. \label{hmaj}
\end{align}
The matrices $M^s_i$ and $M^a_j$ are ``linear independent'' with respect to  the following definition of matrix inner product $\<M_1|M_2\>=V^\dagger_{M_1}\cdot V_{M_2}$, where $V_M$ is the column vector containing all matrix elements of $M$. 
In the following we shall order $\{M^s_i\}$ so that the first $ d $ of them are the $ \G_i $'s in constraint 3. These $\{\G_i\}$ satisfy the Clifford algebra $\{\G_i,\G_j\}=2\delta_{ij}$. Under the above ordering convention,
\be
h(\v k)&&= \sum_{i=1}^{n_s} o_i(\v k) M^s_i + \sum_{j=1}^{n_a} e_j(\v k) M^a_j\nn&&= \sum_{i=1}^{d} o_i(\v k) \G_i + \sum_{i=d+1}^{n_s} o_i(\v k) M_i^s+\sum_{j=1}^{n_a} e_i(\v k) M_i^a. \label{hmaj2}
\ee

\section{The symmetry protection hypothesis}\label{symprot}
Symmetry protection means that under the requirement of \Eq{symm}, there is no non-zero anti-symmetric matrix which anticommutes with all the $\G_i$ in \Eq{effh} and \Eq{hmaj2}.

\section{Spectral Symmetrisation}\label{symmetrise}
Given a $h(\v k)$ satisfying constraints 1-4 in section \ref{lat}, we can create a ``spectral symmetrised" Hamiltonian satisfying the same constraints. \\

To perform spectral symmetrization, we first write $h(\v k)$ in terms of its eigenvalues and eigenvectors
\be
h(\v k) = U_{\v k}^{\dagger} D(\v k) U_{\v k}
\ee
where $ D(\v k) $ is the diagonal matrix formed by the eigenvalues of $h(\v k)$ in descending order. $ U_{\v k} $ contains the eigenvectors. It is the unitary transformation necessary to diagonalize $ h(\v k) $.\\

We first replace the upper and lower halves of the eigenvalues in $ D(\v k) $ by their respective averages. After this replacement, $ D(\v k) $ becomes $ D'(\v k) $ and the  Hamiltonian is given by \be 	h^\prime (\v k ) = U^{\dagger}_{\v{_k}} D'(\v k) U_{\v k}. \label{f0h}\ee Note that in \Eq{f0h} $U_{\v k}$ remains unchanged. From $h^\prime(\v k)$ we  define a new  Hamiltonian by subtracting  the average of the  diagonal element $\bar{E}'(\v k)$ from each element of $D^\prime(\v k)$ so that $D^\prime(\v k)\ra \tilde{D}(\v k)=D^\prime(\v k) -\bar{E}'(\v k) I_n $. Here $I_n$ represents the $2^n \times 2^n$ identity matrix. After the above two steps the Hamiltonian becomes
\be
h(\v k)\ra \tilde{h}(\v k)= U^{\dagger}_{\v{_k}} \tilde{D}(\v k) U_{\v k}.\label{flh}\ee $\tilde{h}(\v k)$ is the ``spectral symmetrised Hamiltonian''. Note that $U_{\v k}$ still remains unchanged. $\tilde{h}(\v k)$ has the important  
property  that
\begin{align}
\tilde{h}(\v k)^2 \propto I_{n} \text{~for~all~}\v k. \label{hflat}
\end{align} 

\begin{figure}[h]
\begin{center}
\includegraphics[scale=0.28]{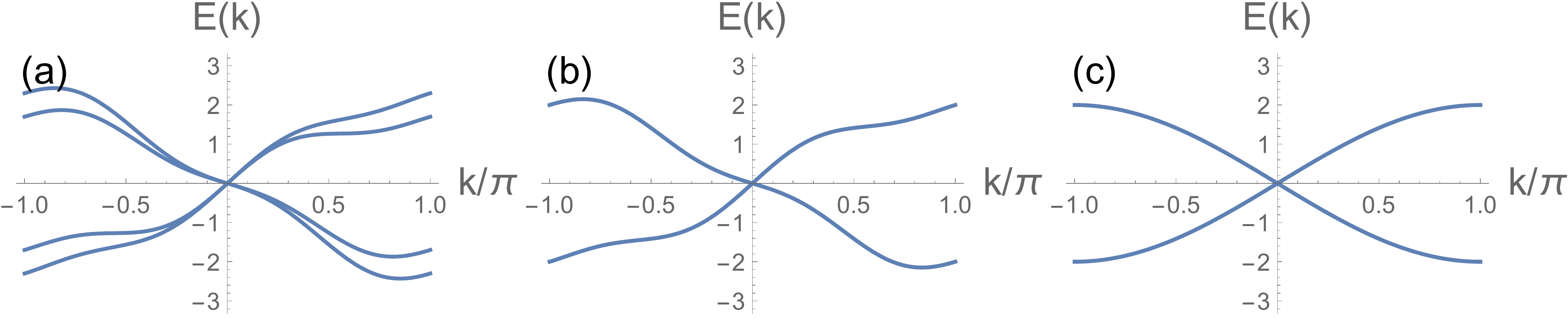}
\caption{From the left to the middle panel we replaced the upper half and lower half of the eigenvalues at each $k$ with their averages. From the middle panel to the right panel we subtracted the average of all eigenvalues from each eigenvalue at each $k$.  }
\label{flat}
\end{center}
\end{figure}

In \ref{preserve}, we show that the spectral symmetrization does not jeopardize constraint 1-4 in section \ref{lat}. In addition,  it preserves the analyticity of $h(\v k)$ in the Brillouin zone region where the energy gap is non-zero. In particular, spectral symmetrization does not affect \Eq{effh}, i.e, \be \tilde{h}(\v k)\ra \sum_{j=1}^d k_j \G_j~~\text{as $\v k\ra 0$.}\label{flh0}\ee 
In \Fig{flat} we show an example of spectral symmetrization in one dimension. \\

The spectral symmetrised $\tilde{h}(\v k)$ can also be written in the form of \Eq{hmaj2}, \ie
\be
\tilde{h}(\v k)&&=\sum_{i=1}^{n_s} \tilde{o}_i(\v k) M_i^s+\sum_{j=1}^{n_a} \tilde{e}_i(\v k) M_i^a\nn&&= \sum_{i=1}^{d} \tilde{o}_i(\v k) \G_i + \sum_{i=d+1}^{n_s} \tilde{o}_i(\v k) M_i^s+\sum_{j=1}^{n_a} \tilde{e}_i(\v k) M_i^a\nn&&:=S(\v k)+A(\v k). \label{hmaj3}
\ee
Here the symmetric matrix  $S(\v k)$ includes the first and the second sums, and the anti-symmetric matrix  $A(\v k)$ includes the third sum.
\\

\section{The Poincar\'e-Hopf Theorem (see, e.g., Ref.\cite{Milnor1965})}\label{PH}
\begin{figure}[h]
\begin{center}
\includegraphics[scale=0.3]{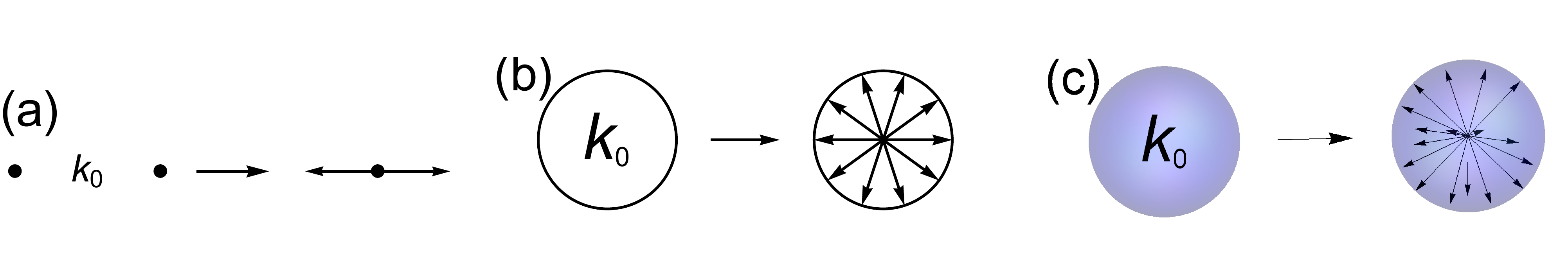}
\caption{The degree 1 maps $\hat{f}(\v k)$ for (a) $d=1$, (b) $d=2$ and (c) $d=3$.}
\label{degree}
\end{center}
\end{figure}

The Poincar\'e-Hopf theorem applies to a $ d $-component vector function $ \v f(\v k)=\{f_1(\v k),...,f_d(\v k)\}$ that vanishes at a discrete set of points $ \{\v k_{n}\} $ on a $ d $-dimensional torus. The theorem states that the ``index" of the $\v k\ra \v f(\v k)$ map at each $ \v k_{n} $ must sum to zero. The meaning of the index is the following.
Pick a closed ball $ D_n $ around each $\v k_n$ so that $ \v k_{n} $ is the only zero of $ \v f(\v k)$ in $ D_n $. We define the index  at $ \v k_{n} $ to be the ``degree" of the map from the boundary of $ D_n $ to the $ (d-1) $-sphere formed by $ \hat{f}(\v k) = \v f (\v k)/|\v f (\v k)| $. For 3D the degree is the Pontryagin index of $ \hat{f}(\v k)$, and in 2D it is the ``winding number" of $ \hat{f}(\v k)$. For 1D the degree is equal to $\left(\hat{f}(k_R)-\hat{f}(k_L)\right)/2$. \Fig{degree} illustrates the degree 1 map for spatial dimension 1,2 and 3.\\

Any zero of $\v f(\v k)$ that has no mapping degree can be removed by infinitesimal changes.  On the other hand, a zero that has non-zero mapping degree can only be shifted but not removed by infinitesimal changes. We assert, without proof, that it is always possible to deform $\tilde{h}(\v k)$ so that  $\tilde{\v o}(\v k):=\{\tilde{o}_1(\v k),....\tilde{o}_d(\v k)\}$ only possess discrete zeros while keeping the symmetrised nature of the energy spectrum. Moreover, around each of the discrete zero $\tilde{\v o}(\v k)$ exhibits a non-zero mapping degree.  \\

Equation \ref{flh0} implies $ \tilde{\v o} (\v k) $ has a degree 1 zero at $\v k=0$. Applying the Poincar\'e-Hopf theorem we conclude that the sum of the mapping degree in the rest of the Brillouin zone must be equal to $ -1 $. Due to the fact that $ \tilde{\v o} (\v k) = - \tilde{\v o} (- \v k) $, and the fact that the degree of mapping is not affected by the simultaneous sign reversal of both  $ \tilde{\v o} $ and $ \v k $, we conclude that the sum of the mapping degree in the Brillouin zone excluding all time-reversal invariant $ \v k $ points must be  an even integer. This, in turn, implies the sum of the mapping degrees across all non-zero time-reversal invariant $ \v k $ points must be an odd integer. (Note that by the oddness of  $\tilde{\v o}(\v k)$, it must vanish at any time reversal invariant $\v k$ point.)
Thus there must exist, at least, one non-zero time-reversal invariant $\v k$ point, say, $\v k_0$, where the mapping degree is an odd integer. 
\\

\section{The reductio ad absurdum proof}\label{prf}

In section \ref{symmetrise} we have shown that given a lattice-regularized $h(\v k)$ satisfying constraints 1-4 in section \ref{lat} there is always a spectral symmetrised $\tilde{h}(\v k)$ which obeys all constraints of $h(\v k)$ and is lattice regularized.\\

In this section we complete the proof of non-regularizability via reductio ad absurdum.
This proof is achieved in two alternative ways. (a) We prove that if $\tilde{\v o}(\v k)$ satisfies a special condition (see below), and if $\tilde{h}(\v k)$ obeys constraints 1-4 of section \ref{lat}, the symmetry-protection  hypothesis must be violated. (b) For $\tilde{\v o}(\v k)$ that violates the special condition, we prove that if $\tilde{h}(\v k)$  satisfies the symmetry-protection hypothesis its energy gap must also close at $\v k_0$. This means constraint 3 of section \ref{lat} is violated. Because proof (b) involves a case-by-case study of all $\hat{T},\hat{Q},\hat{C}$ protected minimal SPN, we leave it to \ref{the proof} and \ref{odd_func}.\\

The symmetries under consideration are generated by the subsets of $\{\hat{T},\hat{Q},\hat{C}\}$. Here 
$$\hat{Q} = i\sum_{\v k\in BZ} ~\chi^T(\v k)~Q~\chi(\v k)$$ is the total charge operator. It  generates the global charge U(1) gauge transformation. $\hat{T}$ and $\hat{C}$ are the generators of  time reversal and charge conjugation symmetries. They act on the fermion operators according to
\be
&&\hat{T}\chi(\v k)\hat{T}^{-1}= T\chi(-\v k)\nn
&&\hat{C}\chi(\v k)\hat{C}^{-1}= C\chi(\v k)
\ee
where $T,Q,C$ are $2^n\times 2^n$ matrices. In \ref{the proof} we list the relevant $T,Q,C$ and all symmetry allowed $\{M_i^s, i=1,...,n_s\}$ and  $\{M_i^a, i=1,...,n_a\}$ in \Eq{hmaj3} for the minimal SPNs in spatial dimensions $1\le d\le 3$.\\

Due to the spectral symmetrization condition, \Eq{hflat}, the $S(\v k)$ and $A(\v k)$ in \Eq{hmaj3} must anticommute. This is because the square of $\tilde{h}(\v k)$ is
\be
\tilde{h}(\v k)^2=S(\v k)^2+A(\v k)^2+\{S(\v k),A(\v k)\},
\ee 
since $\{S(\v k),A(\v k)\}$ is an anti-symmetric matrix, while $\tilde{h}(\v k)^2$ is proportional to the identity matrix, it 
implies 
\be
\{S(\v k),A(\v k)\}=0~~\text{for all}~~ \v k.\label{qq}\ee
Now apply \Eq{qq} to $\v k_0$. Since $\v k_0$ is a time reversal invariant point  $S(\v k_0)=0$, which means $\{\tilde{o}_1(\v k_0),...,\tilde{o}_d(\v k_0)\}=0$.\\

The simplest case to prove the contradiction is when (i) all $d$ functions $\{\tilde{o}_1(\v k_0+\v q),...,\tilde{o}_d(\v k_0+\v q)
\}$ vanish as the same power in $\v q$ as $\v q\ra 0$, and (ii) all other $\tilde{o}_i(\v q)$, namely, $\tilde{o}_{d+1}(\v k_0+\v q),...,\tilde{o}_{n_s}(\v k_0+\v q)$,  vanish as higher power in $\v q$. 
 Under such condition 
 examining $\{S(\v k_0+\v q),A(\v k_0+\v q)\}=0$ to the lowest order in $\v q$ gives us  \be
	\{A(\v k_0),\G_i\}=0, ~~{\rm for}~~i=1,...,d.\label{uy}\ee  
	 \\
	
Equation \ref{uy} implies $A(\v k_0)$ acts like a mass term. Since $A(\v k_0)=\tilde{h}(\v k_0)\ne 0$ (otherwise $\tilde{h}(\v k)$ will have more than one gap node), this violates the symmetry-protection hypothesis. More specifically, including $A(\v k_0)$ in the Hamiltonian
	\be
	H=\int d^d x~\chi^T(\v x)\left[-i\sum_{i=1}^d\G_i\partial_i+A(\v k_0)\right]\chi(\v x)\ee  gaps out the node at $\v k=0$. \\

Under the more general condition, namely when $\{\tilde{o}_1(\v k_0+\v q),...,\tilde{o}_d(\v k_0+\v q)
\}$ do not vanish as the same power in $\v q$, and/or 
   when  $\tilde{o}_{d+1}(\v k_0+\v q),...,\tilde{o}_{n_s}(\v k_0+\v q)$ vanish slower than, or as slowly as, $\tilde{o}_1(\v k_0+\v q),...,\tilde{o}_d(\v k_0+\v q)$ the above proof does not apply.\\
   
     Under such condition we adopt a different proof strategy. Instead, we assume the symmetry-protection hypothesis holds, and show that it is impossible for $\tilde{h}(\v k)$ to have gap node at only a single point in the Brillouin zone. This proof is achieved via a case-by-case study of all
   $\hat{T},\hat{Q},\hat{C}$ symmetry-protected minimal nodal Hamiltonians. Because of the length of the proof we leave it to \ref{the proof} and \ref{odd_func}.

\section{Final discussion: the open issues}\label{dis}
In the preceding discussions we have proven that all minimal nodal Hamiltonians  protected by $\{\hat{T},\hat{Q},\hat{C}\}$ symmetries cannot be regularized on a lattice. Here we list some of the open issues.  The first is the proof for non-minimal symmetry-protected nodal Hamiltonians. Such nodal Hamiltonians can be constructed by stacking the minimal nodal Hamiltonians together. 
Although it is clear that the non-regularizability of the minimal nodal Hamiltonians is a necessary condition for the non-regularizability of non-minimal symmetry-protected nodal Hamiltonians, it remains to be proven that it is a sufficient condition. The second issue concerns the assumption that in the spectral symmetrised Hamiltonian the coefficient functions in front of $\{\G_1,...,\G_d\}$ exhibit isolated zeros. It remains to be proven that it is always possible to deform $\tilde{h}(\v k)$ so that the coefficient functions fulfill such a statement while maintaining the symmetrised spectrum. The third issue is the proof that a general symmetry-protected gapless Hamiltonian can be deformed into the single-node Hamiltonian discussed in this paper. We leave these open issues for future researches.

\appendix

\section{The preservation of constraints 1 to 4 by the spectral symmetrization steps}\label{preserve}
In this section, we show that spectral symmetrization preserves the constraints 1 to 4. As discussed in the main text, if there is anti-unitary symmetry, the spectrum of $h(\v k)$ is symmetric about $E=0$, in which case there is no need for the second step of spectral symmetrization, namely, subtracting the average of eigenenergies.

\subsection{Constraint 1}
The Majorana constraint implies the original Hamiltonian satisfies
\begin{align}
h^{T}(-\v k) &= -h(\v k) \label{maj}
\end{align}
This implies the eigenvalues at $ -\v k $ are the negative of the eigenvalues at $ + \v k $. Thus 
\begin{align}
D(-\v k) = - W^{\dagger}_{\v k} D(\v k) W_{\v k} \label{wdw}
\end{align}
where $ W_{\v k} $ is the unitary transformation necessary to reorder the eigenvalues in $ D(-\v k) $ according to descending order. The Majorana constraint of \Eq{maj} implies
\begin{align*}
U_{-\v k}^T D(-\v k) U_{-\v k}^* = - U_{\v k}^{\dagger} D(\v k) U_{\v k}.
\end{align*}
We substitute \Eq{wdw} into the above equation,
\begin{align}
&&U_{-\v k}^T W^{\dagger}_{\v k} D(\v k) W_{\v k} U_{-\v k}^* = U_{\v k}^{\dagger} D(\v k) U_{\v k}\nn
&&\Rightarrow  U_{\v k} U_{-\v k}^T W^{\dagger}_{\v k} D(\v k) W_{\v k} U_{-\v k}^* U_{\v k}^{\dagger} =  D(\v k) \label{wtdwt}
\end{align}
The second line of the above equation can be rewritten as 
\be
Z_{\v k} D(\v k)Z^\dagger_{\v k}= D(\v k),
\label{zdz}\ee
where the unitary matrix $ Z_{\v k} = U_{\v k} U_{-\v k}^T W^{\dagger}_{\v k} $. In order for \Eq{zdz} to hold, $ Z_{\v k} $ needs to be block diagonalized where each block is spanned by the degenerate eigenvectors of $ D(\v k) $. Within each block, $ D(\v k) $ is proportional to an identity matrix. \\

After the first step of spectral symmetrization $D(\v k)\ra D'(\v k)$. Since $ D'(\v k) $ is still proportional to the same identity matrix in each block of $ D(\v k) $, it follows that conjugation by $ Z(\v k) $ still leaves it invariant, i.e.,
\be
Z_{\v k} D'(\v k) Z^\dagger_{\v k}=U_{\v k} U_{-\v k}^T W^{\dagger}_{\v k} D'(\v k) W_{\v k} U_{-\v k}^* U_{\v k}^{\dagger} &=  D'(\v k).
\label{wrev}\ee
Given \Eq{wrev} we can multiply the unitary matrices in the reverse order to arrive at
\begin{align*}
U_{-\v k}^T D'(-\v k) U_{-\v k}^* = - U_{\v k}^{\dagger} D'(\v k) U_{\v k},
\end{align*}
which means
\be
{h'}^{T}(-\v k) &= -h'(\v k).\label{fmj}
\ee

Equation \ref{fmj} implies that the spectrum of $h'(\v k)$ flips sign upon the reversal of $\v k$. As a result, the average of the diagonal elements $\bar{E'}(\v k)$ subtracted in the second step of spectral symmetrization,  obeys 
\be
\bar{E'}(-\v k)=-\bar{E'}(\v k).\ee  Consequently the subtracted piece  $\bar{E'}(\v k)I_n$ obeys the Majorana constraint, i.e.,
\be
\left(\bar{E'}(-\v k)I_n\right)^T=-\bar{E'}(\v k)I_n.
\ee
This means if $h'(\v k)$ satisfies the Majorana constraint, so does $\tilde{h}(\v k)$ after the subtraction.

\subsection{Constraint 2}\label{preserve2}
The periodicity constraint is given by
\begin{align}
h(\v k) = h(\v k + \v G) \label{period}
\end{align}\
where $ \v G $ is any reciprocal lattice vector. This means
\begin{align*}
U_{\v k}^{\dagger} D(\v k) U_{\v k} &= U_{\v k+ \v G}^{\dagger} D(\v k+ \v G) U_{\v k+ \v G}\\
\Rightarrow  D(\v k+ \v G)  &= U_{\v k+ \v G} U_{\v k}^{\dagger} D(\v k) U_{\v k} U_{\v k+ \v G}^{\dagger}.
\end{align*}
Since $ D(\v k+ \v G) = D(\v k) $ (periodicity in Hamiltonian implies periodicity in the eigenvalues), we have
\begin{align}
D(\v k)  &= U_{\v k+ \v G} U_{\v k}^{\dagger} D(\v k) U_{\v k} U_{\v k+ \v G}^{\dagger} \equiv Y_{\v k}D(\v k)Y^\dagger_{\v k}. \label{wtdwt2}
\end{align}
Here the unitary matrix $ Y_{\v k} =  U_{\v k+ \v G} U_{\v k}^{\dagger} $ needs to be block diagonalized where each block is spanned by the degenerate eigenvectors of $ D(\v k) $. Within each block, $ D(\v k) $ is proportional to an identity matrix. Since $ D'(\v k) $ is still proportional to the same identity matrix in each block of $ D(\v k) $, conjugation by $ Y(\v k) $  leaves $ D'(\v k) $ invariant. Thus
\begin{align*}
D'(\v k)  &= U_{\v k+ \v G} U_{\v k}^{\dagger} D'(\v k) U_{\v k} U_{\v k+ \v G}^{\dagger}.
\end{align*}
Since $ D'(\v k) $ on the LHS equals to $ D'(\v k + \v G) $, it follows that 
\begin{align*}
D'(\v k+\v G)  &= U_{\v k+ \v G} U_{\v k}^{\dagger} D'(\v k) U_{\v k} U_{\v k+ \v G}^{\dagger}.
\end{align*}
Multiplying the unitary matrices in reverse order leads to
\begin{align*}
h'(\v k) = h'(\v k + \v G).
\end{align*}

Because $h'(\v k)$ satisfies the periodicity constraints, so does the average of its eigenvalues $\bar{E'}(\v k)$. Hence $\bar{E'}(\v k)I_n$, subtracted in the second step of spectral symmetrization, obeys the Brillouin zone periodicity. As as result, $\tilde{h}(\v k)$ satisfies the periodicity constraint. \\

Another important part of constraint 2 is the analytic nature of $h(\v k)$. In the following we show that in the $\v k$ region where the spectrum is gapped, spectral symmetrization does not spoil analyticity.\\

Let's consider shifting $\v k$ to $\v k+\e\hat{n}$, where 
$\hat{n}$ is an unit vector and $\e$ is an infinitesimal. Under such infinitesimal shift

\be
h(\v k)\ra h(\v k+ \epsilon  \hat{\v n}).\label{conti}
\ee
In the mathematics literature, e.g., theorem 1 in Chapter I (page 42) of Ref.\cite{Rellich1969}), there is the following theorem.

\begin{blockquote}\
{\bf Theorem} Let $h(x)$ be a finite dimensional Hermitian matrix function of a parameter $x$. If the polynomial expansion of $h(x)$ around $x=0$ has a finite radius of convergence (i.e. analytic), then there exists a basis in which both the eigenvalues and the orthonormal set of eigenvectors of $h(x)$ have a convergent power series expansion within the same radius.
\end{blockquote}
It is important to note that this theorem applies whether there are degeneracies in the eigenvalues of $h(0)$ or not.\\

\begin{figure}[h]
	\begin{center}
		\includegraphics[scale=0.6]{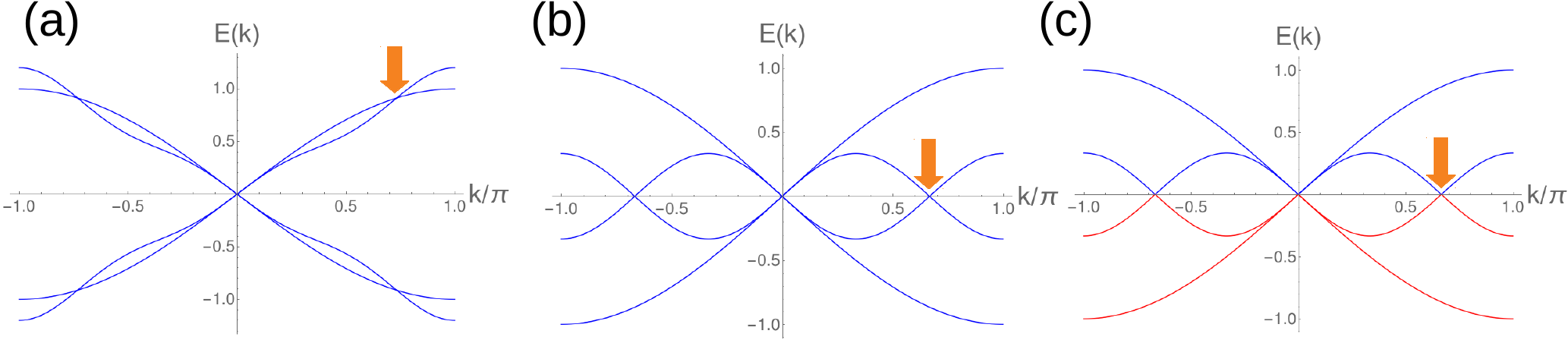}
		\caption{Examples of band crossing at $ k\neq 0$ in 1D. (a) The orange arrow points at a $k$ point where  band crossing occurs while the energy gap remains non-zero. (b) The orange arrow points at a gap-closing $k$ point. (c) After the energy eigenvalues are sorted into upper (blue) and lower (red) halves, the eigenvalues and eigenvectors are no longer analytic across the gap-closing $k$ point.}
		\label{gapped_gapless_k}
	\end{center}
\end{figure}

Applying this theorem to our problem, the analytic nature of $h(\v k)$ around $\v k$ implies the existence of a basis in which  the eigenvalues and eigenvectors of $h(\v k+ \epsilon  \hat{\v n})$ is an analytic function of $\e$ in any direction $\hat{n}$.
This means we can choose a basis so that both $D(\v k+ \epsilon  \hat{\v n})$ and $U (\v k+ \epsilon  \hat{\v n})$ in 
\be
h(\v k+ \epsilon  \hat{\v n}) = U^{\dagger}(\v k + \epsilon  \hat{\v n}) 
D(\v k+ \epsilon  \hat{\v n}) U (\v k+ \epsilon  \hat{\v n})
\ee
are analytic functions of $\epsilon$. In regions where $h(\v k)$ is gapped we can sort the eigenvalues into an upper half and a lower half so that no interchange of eigenvalues between the two parts take place as $\v k$ moves around. Note that this is also true when there is crossing between the bands in the upper or lower halves (see Figure \ref{gapped_gapless_k} (a)).\\

Under such (no gap closure) condition, the spectral symmetrization does not change the analyticity of the Hamiltonian, because the eigenvectors are unchanged and the average of the upper/lower half of the eigenvalues as well as the average of all eigenvalues are analytic in $\v k$.  This is no longer true when $\v k$ moves across a gap closing point (see Figure \ref{gapped_gapless_k} (b)). In that case there exists eigenvalues (and their associated eigenvectors) that move from the upper to the lower part (and vice versa). Under this condition although the original eigenvalues and eigenvectors are analytic in $\v k$, {\it the sorted ones are not} (see Figure \ref{gapped_gapless_k} (c)).\\

Thus if $h$ is analytic and gapped in a neighborhood of $\v k$ the spectral symmetrised 
$\tilde{h}$ is analytic too. In contrast, spectral symmetrization does not maintain the analytic nature the Hamiltonian if $\v k$ moves across gap nodes.

\subsection{Constraint 3}

The spectral symmetrization step clearly does not collapse the energy gap. \\

At $\v k=0$, since all eigen-energies are zero, no spectral symmetrization is necessary. Moreover, the spectral symmetrization does not change the fact that $h(\v k)\ra \sum_j k_j\G_j$ as $\v k\ra 0$, because $\sum_j k_i\G_j$ already satisfies the spectral symmetrization condition. Together, the above arguments imply that spectral symmetrization preserves constraint 3.

\subsection{Constraint 4}
\subsubsection{The unitary symmetries}
The unitary symmetries require
\begin{align*}
U_{\b}^{\dagger} h(\v k) U_{\b} = h(\v k),
\end{align*}
which means
\begin{align}
&U_{\b}^{\dagger} U_{\v k}^{\dagger} D(\v k) U_{\v k} U_{\b} = U_{\v k}^{\dagger} D(\v k) U_{\v k} \nn
&\Rightarrow Q_{\v k}D(\v k)Q^\dagger_{\v k}\equiv U_{\v k}U_{\b}^{\dagger} U_{\v k}^{\dagger} D(\v k) U_{\v k} U_{\b} U_{\v k}^{\dagger}=  D(\v k).
\end{align}
Again, the unitary matrix $ Q_{\v k} =  U_{\v k}U_{\b}^{\dagger} U_{\v k}^{\dagger} $ needs to be block diagonalized where each block is spanned by the degenerate eigenvectors of $ D(\v k) $. Within each block, $ D(\v k) $ is proportional to an identity matrix.\\

After spectral symmetrization, $ D'(\v k) $ is still proportional to the same identity matrix in each block of $ D(\v k) $, hence conjugation by $ Q(\v k) $  leaves $ D'(\v k) $ invariant. Thus
\begin{align*}
U_{\v k}U_{\b}^{\dagger} U_{\v k}^{\dagger} D'(\v k) U_{\v k} U_{\b} U_{\v k}^{\dagger}&=  D'(\v k).
\end{align*}

Multiplying the unitary matrices in reverse order leads to
\begin{align*}
U_{\b}^{\dagger} h'(\v k) U_{\b} =h'(\v k).
\end{align*}\\

The $\bar{E}'(\v k)I_n$, subtracted in the second step of spectral symmetrization, clearly satisfies the unitary symmetry constraint, namely, 
\begin{align*}
U_{\b}^{\dagger}\left(\bar{E}'(\v k)I_n\right) U_{\b} = \bar{E}'(\v k)I_n.
\end{align*}
As a result, the subtraction does not jeopardize the unitary symmetry.

\subsubsection{The anti-unitary symmetries}
The anti-unitary symmetries require
\begin{align*}
A_{\a}^{\dagger} h(-\v k )^{*} A_{\a} = h(\v k).
\end{align*}
The Majorana constraint \Eq{maj} converts the above equation to
\begin{align}
-A_{\a}^{\dagger} h(\v k) A_{\a} = h(\v k). \label{aha}
\end{align}
Among other things, this means the eigenvalues of $ h(\v k) $ are in $ \pm $ pairs, which means 
\begin{align}
D(\v k) = X_{\v k}^{\dagger} \left(-D(\v k)\right) X_{\v k}. \label{xdx}
\end{align}
Where $ X_{\v k} $ is a unitary matrix necessary to reorder the eigenvalues of $ -D(\v k) $ in descending order. Equation \ref{aha} implies
\be
&A_{\a}^{\dagger}U_{\v k}^{\dagger} D(\v k) U_{\v k} A_{\a} =  -U_{\v k}^{\dagger} D(\v k) U_{\v k}\nn
\Rightarrow &U_{\v k}A_{\a}^{\dagger}  U_{\v k}^{\dagger} D(\v k) U_{\v k} A_{\a} U_{\v k}^{\dagger} =  - D(\v k) 
\ee
Now we use \Eq{xdx} to convert the last line of the above equation to
\begin{align*}
O_{\v k}D(\v k)O^\dagger_{\v k}\equiv U_{\v k}A_{\a}^{\dagger}  U_{\v k}^{\dagger} X_{\v k}^{\dagger} D(\v k) X_{\v k} U_{\v k} A_{\a} U_{\v k}^{\dagger} &=  D(\v k) 
\end{align*}
Like before, the unitary matrix $ O_{\v k} =  U_{\v k}A_{\a}^{\dagger}  U_{\v k}^{\dagger} X_{\v k}^{\dagger} $ needs to be block diagonalized where each block is spanned by degenerate eigenvectors of $ D(\v k) $.  Within each block, $ D(\v k) $ is proportional to an identity matrix.\\

After spectral symmetrization $ D'(\v k) $ is still proportional to the same identity matrix in each block of $ D(\v k) $, hence conjugation by $ O(\v k) $  leaves $ D'(\v k) $ invariant, i.e.,
\begin{align*}
U_{\v k}A_{\a}^{\dagger}  U_{\v k}^{\dagger} X_{\v k}^{\dagger} D'(\v k) X_{\v k} U_{\v k} A_{\a} U_{\v k}^{\dagger} &=  D'(\v k).
\end{align*}
Since the same $ X_{\v k} $ can reverse the ordering of eigenvalues in $ D'(\v k) $, the above equation turns into
\begin{align*}
U_{\v k}A_{\a}^{\dagger}  U_{\v k}^{\dagger} \left(- D'(\v k)\right) U_{\v k} A_{\a} U_{\v k}^{\dagger} &=  D'(\v k).
\end{align*}

Multiplying the unitary matrices in reverse order leads to
\be
-A_{\a}^{\dagger} h'(\v k) A_{\a} =h'(\v k).\label{anti}
\ee
Since \Eq{anti} implies the eigenvalues of $h'(\v k)$ are symmetric with respect to $E=0$, there is no subtraction step 
needed. Hence $\tilde{h} (\v k)=h'(\v k)$, and  
\begin{align*}
-A_{\a}^{\dagger} \tilde{h}(\v k) A_{\a} =\tilde{h}(\v k)\Rightarrow A_{\a}^{\dagger} \tilde{h}(-\v k )^{*} A_{\a} = \tilde{h}(\v k).
\end{align*}

\section{Under the symmetry-protection hypothesis it is impossible for the gap of $\tilde{h}(\v k)$ to close at only a single point in the Brillouin zone}\label{the proof}
In this appendix, we prove that the symmetry protection constraint plus constraints 1,2,4  and \Eq{effh} in section \ref{lat}  lead to the violation of the single gap node assumption in constraint 3. More specifically, we prove that under the conditions described above $A(\v k_0)=0$. Here $A(\v k)$ is defined in \Eq{hmaj3} and $\v k_0$ is the non-zero time reversal invariant point  discussed in the main text. Since $S(\v k_0)=0$ this implies $\tilde{h}(\v k_0)=0$. This violates the statement that energy gap  closes only at $\v k=0$. This proof addresses the generic situations discussed in section \ref{prf} of the main text.\\

For each symmetry group generated by a subset of $\{\hat{T},\hat{Q},\hat{C}\}$ we focus on the 
minimal models where the number of components, $n_0$, in $\chi(\v k)$ is the minimum. 
This is the minimal number of components required to realize a particular SPN. Under this condition the dimension of all associated matrices is $n_0\times n_0$.
In other words, $n_0$ is the minimum integer for which there exists $n_0$-by-$n_0$ matrices representing the available symmetries and $\Gamma_1,...,\G_d$ ($d$ is the spatial dimension). Here the $\{\G_i\}$ obey the symmetry requirement (constraint 4 in section \ref{lat}) and satisfy the Clifford algebra $\{\G_i,\G_j\}=2\delta_{ij}$.\\

For each spatial dimension $d$, we will go through all the symmetry groups $G$ which gives rise to an SPN. (These groups  protect non-trivial SPT's in $d+1$ dimensions.) For each $(d,G)$ we write down the number $n_0$, the symmetry matrices, and the most general form of $S(\v k)$ and $A(\v k)$ allowed by symmetry.\\

To characterize each symmetry group we shall  use the short hand $$G^\pm([~]_\pm,[~]_\pm,[~]_\pm).$$  Between the square brackets we insert $T,Q$ or $C$ (the maximal number of symbols in the argument of $G$ is 3). The subscript of the symbols, when present, denotes whether the matrix representing the ${\hat{T},\hat{Q},\hat{C}}$ squares to identity or minus identity. The superscript on $G$ specifies whether the time reversal matrix $T$ commutes ($+$) or anticommutes ($-$) with the charge conjugation  matrix $C$. The matrix $Q$ always anticommutes with $T$ and $C$, and always squares to minus identity. Hence we do not bother to attach a subscript to $Q$, nor do we need to specify the commutator between $Q$ and $T,C$. To simplify the notation we shall abbreviate the Pauli matrices $\s_0,\s_x, \s_y, \s_z, i \s_y$ as $I,X,Y,Z,E$, respectively. When two Pauli matrices appear next to each other it means tensor product. For example $EX$ means $i\s_y\otimes\s_x$.\\

The proof is based on the following facts.
\begin{enumerate}
\item After spectral symmetrization, the Hamiltonian is given by \Eq{hmaj3} in the main text, where 
 $\{ S(\v k), A(\v k)\}=0$ and $S(\v k)^2 + A(\v k)^2 \propto I_n$.
\item As shown in \ref{preserve2} the spectral symmetrization preserves the analytic nature of $h(\v k)$ in regions of $\v k$ where the spectrum of $h(\v k)$ is fully  gapped. Hence in the gapped region of $h(\v k)$, $\tilde{h}(\v k)$ and the coefficient functions $\tilde{o}_i (\v k)$ and $\tilde{e}_j(\v k)$ in \Eq{hmaj3} are analytic.
\item The Poincar\'e -Hopf theorem implies the mapping degree of $$\{ \tilde{o}_1 (\v k),\tilde{o}_2 (\v k) ,... ,\tilde{o}_d (\v k) \}$$ is odd around, at least, one other time-reversal invariant point $\v k_0 \neq 0 $.
\end{enumerate}

In addition, for the ease of later discussions, we define the curves $\{\mathcal{C}_i,~i=1,...,d\}$  near $\v k_0$  as follows.
\begin{define}
Given $i\in\{1,\dots,d\}$, let's consider the map $\v q \rightarrow (\tilde{o}_1(\v k_0 + \v q) ,\dots, \tilde{o}_d(\v k_0 + \v q))$ from any circle of radius $| \v q|=r>0$. Due to the non-zero degree of this map there must exist, at least, one point $\v q$ on the circle such that $\tilde{o}_j(\v k_0+\v q) = 0 $ for $j\neq i$ and $\tilde{o}_i(\v k_0+\v q) > 0 $. Let's select such a point. Because the coefficient functions are continuous we can connect the points for different $r$ into a curve $\mathcal{C}_i$ which approaches the point $\v k_0$ as $r\ra 0$. 
\end{define}\label{curve}

\subsection{{\bf 1D SPNs}}
\vspace{0.1 in}

\subsubsection{$G(\emptyset)$, {\rm or equivalently} $G(C_+)$  {\rm after block-diagonalizing} $C$}

\begin{align*}
G(\emptyset),~~n_0=1,~~
\begin{array}{c| c} S(k) & \tilde{o}_1(k)   \\ \hline  A(k) & 0     \end{array} 
\end{align*}

As mentioned in the main text, this is a chiral SPN. It is not regularizable because the continuity and the Brillouin zone periodicity contradict with each other.\\

\subsubsection{$G(T_-)$, {\rm or equivalently} $G^+(T_-,C_+)$ {\rm after block-diagonalizing} $C$}

\begin{align*}
 T=E,
~~n_0=2,~~
\begin{array}{c| c} S(k) & \tilde{o}_1(k ) X + \tilde{o}_2 (k) Z  \\ \hline  A(k) & 0     \end{array} 
\end{align*}
Since there is no $A(k) \implies A(k_0)=0$.\\

\subsubsection{$G(C_-)$, {\rm or equivalently} $G(Q)$ {\rm after identifying}  $C$ {\rm with} $Q$}

\begin{align*}
C=E,
~~n_0=2,~~
\begin{array}{c| c} S(k) & \tilde{o}_1(k) I  \\ \hline  A(k) & \tilde{e}_1 (k)Y     \end{array} 
\end{align*}

Since $S(\v k)\propto I$ this is a chiral SPN. It is not regularizable because the continuity and the Brillouin zone periodicity contradict with each other.\\

\subsubsection{$G^-(T_+, C_+)$}

\begin{align*}
T=Z, C=X,~~n_0=2,~~
\begin{array}{c| c} S(k) & \tilde{o}_1 (k) X  \\ \hline  A(k) & 0     \end{array} 
\end{align*}
Since there is no $A(k)$ $\implies A(k_0)=0$.

\subsubsection{$G^-(T_-, C_+)$}

\begin{align*}
T=E, C=Z,~~,n_0=2,~~
\begin{array}{c| c} S(k) & \tilde{o}_1 (k) Z  \\ \hline  A(k) & 0     \end{array} 
\end{align*}
Since there is no $A(k)$ $\implies A(k_0)=0$.

\subsubsection{$G^-(T_-, C_-)$, {\rm or equivalently} $G(Q,T_-)$ {\rm after identifying}  $C$ {\rm with} $Q$}

\begin{align*}
T=ZE, C=EI,~~n_0=4,~~
\begin{array}{c| c} S(k) & \tilde{o}_1 (k) YY + \tilde{o}_2 (k) IX +\tilde{o}_3 (k) IZ  \\ \hline  A(k) &  \tilde{e}_1 (k) YI    \end{array} 
\end{align*}
$\{ S(k) , A(k) \}=0 $ implies

\begin{align*}
\begin{cases}
\tilde{o}_1(k) \tilde{e}_1(k) = 0 \\
\tilde{o}_2(k) \tilde{e}_1(k) = 0 \\
\tilde{o}_3(k) \tilde{e}_1(k) = 0 \\
\end{cases}
\end{align*}

Because the mapping degree of $\tilde{o}_1(k)$ is odd in the neighborhood of $k = k_0$, it requires $\tilde{o}_1(k)$ to be non-zero when $k$ is in the neighborhood but not equal to $k_0$. This implies $\tilde{e}_1(k)=0$ in the neighborhood of $k_0$. The continuity of $\tilde{e}_1(k)$ implies $\tilde{e}_1(k_0)=0$, which in turn implies $A(k_0)=0$.

\subsubsection{$G(Q, C_+)$}

\begin{align*}
Q=E, C=Z,~~n_0=2,~~
\begin{array}{c| c} S(k) &  \tilde{o}_1(k) I \\ \hline A(k) & 0    \end{array} 
\end{align*}

Since $S(\v k)\propto I$ this is a chiral SPN. It is not regularizable because the continuity and the Brillouin zone periodicity contradict with each other.\\

\subsubsection{$G(Q, C_-)$}

\begin{align*}
Q=EI, C=ZE,~~n_0=4,~~
\begin{array}{c| c} S(k) & \tilde{o}_1(k) II   \\ \hline A(k) & \tilde{e}_1 (k)YX + \tilde{e}_2 (k) YZ + \tilde{e}_3 (k)IY    \end{array} 
\end{align*}

Since $S(\v k)\propto II$ this is a chiral SPN. It is not regularizable because the continuity and the Brillouin zone periodicity contradict with each other.\\

\subsubsection{$G^+(Q,T_-, C_+)$ or equivalently $G^-(Q,T_-, C_+)$ after identifying $C$ with $QC$}

\begin{align*}
Q=EI, T=ZE ,C=XX,~~n_0=4,~~
\begin{array}{c| c} S(k) & \tilde{o}_1 (k)YY+ \tilde{o}_2 (k)IX  \\ \hline A(k) & 0    \end{array} 
\end{align*}
Since there is no $A(k)$ $\implies A(k_0)=0$.

\subsection{{\bf 2D SPNs}}
\vspace{0.1 in}

\subsubsection{$G(T_-)$, {\rm or equivalently} $G^+(T_-, C_+)$  {\rm after block-diagonalizing} $C$}
\begin{align*}
T=E,~~n_0=2,~~
\begin{array}{c| c} S(\v k) & \tilde{o}_1 (\v k) X + \tilde{o}_2 (\v k)Z  \\ \hline A(\v k) & 0    \end{array} 
\end{align*}
Since there is no $A(k)$ $\implies A(k_0)=0$.

\subsubsection{$G^+(T_+,C_-)$ or equivalently, $G^+(T_-,C_-)$ after identifying $T_-$ with $T_+C_-$}

\begin{align*}
T=ZI, C=ZE,~~n_0=4,~~
\begin{array}{c| c} S(\v k) & \tilde{o}_1 (\v k) XX + \tilde{o}_2 (\v k) XZ  \\ \hline A(\v k) & \tilde{e}_1 (\v k) YX + \tilde{e}_2 (\v k)YZ    \end{array} 
\end{align*}
Here $\{ S(\v k) , A(\v k) \}=0 $ implies
\begin{align*}
\tilde{o}_2(\v k) \tilde{e}_1(\v k) = \tilde{o}_1(\v k) \tilde{e}_2(\v k) \\
\end{align*}
We examine the above equation in the neighborhood of $\v k_0$ by expanding $\v k  = \v k_0 + \v q$. \\

On the curve $\mathcal{C}_1$ defined in \ref{curve} with $d=2$, for any $r=|\v q|\neq 0$,
\be
0 = \tilde{o}_2(\v k) \tilde{e}_1(\v k) = \tilde{o}_1(\v k) \tilde{e}_2(\v k) 
\ee
Because $\tilde{o}_1(\v k)>0$ it implies $\tilde{e}_2(\v k)=0$. By the continuity of $\tilde{e}_2(\v k)$ we conclude $\tilde{e}_2(\v k)=0$ at $r=0$. In other words $\tilde{e}_2(\v k_0)=0$. We can repeat this argument by looking at $\mathcal{C}_2$. This will lead to  $\tilde{e}_1(\v k_0)=0$. Combining the above results, we obtain  $A(\v k_0)=0$.

\subsubsection{$G^-(T_-,C_-)$, {\rm or equivalently} $G(Q,T_-)$ {\rm after identifying}  $C$ {\rm with} $Q$}

\begin{align*}
T=ZE, C=EI,~~n_0=4,~~
\begin{array}{c| c} S(\v k) & \tilde{o}_1 (\v k)YY + \tilde{o}_2 (\v k) IX + \tilde{o}_3 (\v k) IZ  \\ \hline A(\v k) & \tilde{e}_1 (\v k)YI    \end{array} 
\end{align*}
Here $\{ S(\v k) , A(\v k) \}=0 $ implies

\begin{align*}
\begin{cases}
\tilde{o}_1(\v k) \tilde{e}_1(\v k) = 0 \\
\tilde{o}_2(\v k) \tilde{e}_1(\v k) = 0 \\
\tilde{o}_3(\v k) \tilde{e}_1(\v k) = 0 \\
\end{cases}
\end{align*}
Let's focus on the first two equations. 
\\

On the curve $\mathcal{C}_1$ defined in \ref{curve} with $d=2$, for any $r=|\v q|\neq 0$,
\be
\tilde{o}_1(\v k) \tilde{e}_1(\v k) = 0
\ee
Because $\tilde{o}_1(\v k)>0$ it implies $\tilde{e}_1(\v k)=0$. By the continuity of $\tilde{e}_1(\v k)$ we conclude $\tilde{e}_1(\v k)=0$ at $r=0$. In other words $\tilde{e}_1(\v k_0)=0$. This means $A(\v k_0)=0$.

\subsubsection{$G^+(Q,T_+,C_-)$, {\rm or equivalently} $G^-(Q,T_+,C_-)$ {\rm after identifying}  $C$ {\rm with} $QC$}

\begin{align*}
Q=EII, T=ZII, C=ZEI,~~n_0=8,~~
\begin{array}{c| c} S(\v k) & \tilde{o}_1 (\v k)YXY + \tilde{o}_2 (\v k)YZY  \\ \hline A(\v k) & \tilde{e}_1 (\v k)YXX + \tilde{e}_2 (\v k)YXZ + \tilde{e}_3 (\v k)YXI +
\\ & \tilde{e}_4 (\v k)YZX + \tilde{e}_5 (\v k)YZZ + \tilde{e}_6 (\v k)YZI    \end{array} 
\end{align*}
Here $\{ S(\v k) , A(\v k) \}=0 $ implies

\begin{align*}
\begin{cases}
\tilde{o}_1(\v k) \tilde{e}_5(\v k)-\tilde{o}_2(\v k) \tilde{e}_2(\v k) = 0 \\
\tilde{o}_1(\v k) \tilde{e}_4(\v k)+\tilde{o}_2(\v k) \tilde{e}_1(\v k) = 0 \\
\tilde{o}_1(\v k) \tilde{e}_3(\v k)+\tilde{o}_2(\v k) \tilde{e}_6(\v k) = 0 \\
\end{cases}
\end{align*}
We examine the above equation in the neighborhood of $\v k_0$ by expanding $\v k  = \v k_0 + \v q$.
On the curve $\mathcal{C}_1$ defined in \ref{curve} with $d=2$, for any $r=|\v q|\neq 0$,
\begin{align*}
&\begin{cases}
\tilde{o}_1(\v k) \tilde{e}_5(\v k) = 0 \\
\tilde{o}_1(\v k) \tilde{e}_4(\v k) = 0 \\
\tilde{o}_1(\v k) \tilde{e}_3(\v k)= 0 \\
\end{cases}
\end{align*}
Because $\tilde{o}_1(\v k)>0$ it implies 
$\tilde{e}_5(\v k) =\tilde{e}_4(\v k)= \tilde{e}_3(\v k)= 0$.  By the continuity of $\tilde{e}_{3,4,5}(\v k)$ we conclude $\tilde{e}_5(\v k) =\tilde{e}_4(\v k)= \tilde{e}_3(\v k)= 0$ at $r=0$, or in other words, $\tilde{e}_5(\v k_0) =\tilde{e}_4(\v k_0)= \tilde{e}_3(\v k_0)= 0$. 
We can repeat this argument by looking at 
$\mathcal{C}_2$, which will lead to  $\tilde{e}_1(\v k_0)=\tilde{e}_2(\v k_0)=\tilde{e}_6(\v k_0)=0$. Combining these results we conclude $A(\v k_0)=0$.

\subsubsection{$G^+(Q,T_-,C_+)$, {\rm or equivalently} $G^-(Q,T_-,C_+)$ {\rm after identifying}  $C$ {\rm with} $QC$}

\begin{align*}
Q=EI, T=ZE, C=ZI,~~n_0=4,~~
\begin{array}{c| c} S(\v k) & \tilde{o}_1 (\v k)IX + \tilde{o}_2 (\v k)IZ  \\ \hline A(\v k) & 0    \end{array} 
\end{align*}
Since there is no $A(k)$ $\implies A(k_0)=0$.

\subsubsection{$G^+(Q,T_-,C_-)$, {\rm or equivalently} $G^-(Q,T_-,C_-)$ {\rm after identifying}  $C$ {\rm with} $QC$}

\begin{align*}
&Q=EII, T=ZEI, C=ZIE,~~n_0=8,\nn
&\begin{array}{c| c} S(\v k) & \tilde{o}_1 (\v k)YYX + \tilde{o}_2 (\v k)YYZ + \tilde{o}_3 (\v k)IXI + \tilde{o}_4 (\v k)IZI  \\ \hline A(\v k) &\tilde{e}_1(\v k) YIX+ \tilde{e}_2 (\v k)YIZ+ \tilde{e}_3 (\v k)IXY + \tilde{e}_4 (\v k)IZY   \end{array} 
\end{align*}
Here $\{ S(\v k) , A(\v k) \}=0 $ implies

\begin{align*}
\begin{cases}
\tilde{o}_1(\v k) \tilde{e}_1(\v k)+\tilde{o}_2(\v k) \tilde{e}_4(\v k) = 0 \\
\tilde{o}_1(\v k) \tilde{e}_3(\v k)+\tilde{o}_2(\v k) \tilde{e}_2(\v k) = 0 \\
\tilde{o}_1(\v k) \tilde{e}_1(\v k)+\tilde{o}_2(\v k) \tilde{e}_2(\v k) = 0 \\
\tilde{o}_3(\v k) \tilde{e}_2(\v k)-\tilde{o}_1(\v k) \tilde{e}_4(\v k) = 0 \\
\tilde{o}_4(\v k) \tilde{e}_1(\v k)-\tilde{o}_2(\v k) \tilde{e}_3(\v k) = 0 \\
\tilde{o}_3(\v k) \tilde{e}_3(\v k)+\tilde{o}_4(\v k) \tilde{e}_4(\v k) = 0 \\
\end{cases}
\end{align*}
Let's focus on the first three equations. We examine these equations in the neighborhood of $\v k_0$ by expanding $\v k  = \v k_0 + \v q$. On the curve $\mathcal{C}_1$ defined in \ref{curve} with $d=2$, for any $r=|\v q|\neq 0$,
\begin{align*}
&\begin{cases}
\tilde{o}_1(\v k) \tilde{e}_1(\v k) = 0 \\
\tilde{o}_1(\v k) \tilde{e}_3(\v k) = 0 \\
\tilde{o}_1(\v k) \tilde{e}_1(\v k) = 0 \\
\end{cases}
\end{align*}
Because $\tilde{o}_1(\v k)>0$ it implies $\tilde{e}_1(\v k)= \tilde{e}_3(\v k)=0$. By the continuity of $\tilde{e}_{1,3}(\v k)$ we conclude $\tilde{e}_1(\v k)= \tilde{e}_3(\v k)=0$ at $r=0$. In other words $\tilde{e}_1(\v k_0)= \tilde{e}_3(\v k_0)=0$. We can repeat this argument by looking at 
$\mathcal{C}_2$, which will lead to  $\tilde{e}_2(\v k_0)=\tilde{e}_4(\v k_0)=0$. Combining these results we conclude $A(\v k_0)=0$.

\subsection{{\bf 3D SPNs}}
\vspace{0.1 in}

\subsubsection{$G(C_-)$, {\rm or equivalently} $G(Q)$ {\rm after identifying}  $C$ {\rm with} $Q$}

\begin{align*}
C=EI,~~n_0=4,~~
\begin{array}{c| c} S(\v k) & \tilde{o}_1 (\v k)YY + \tilde{o}_2 (\v k)IX + \tilde{o}_3 (\v k)IZ  \\ \hline  A(\v k) & \tilde{e}_1 (\v k)YX + \tilde{e}_2 (\v k)YZ + \tilde{e}_3 (\v k)YI + \tilde{e}_4 (\v k)IY     \end{array} 
\end{align*}
Here $\{ S(\v k) , A(\v k) \}=0 $ implies

\begin{align*}
\begin{cases}
\tilde{o}_2(\v k) \tilde{e}_3(\v k) = 0 \\
\tilde{o}_3(\v k) \tilde{e}_3(\v k) = 0 \\
\tilde{o}_1(\v k) \tilde{e}_3(\v k) = 0 \\
\tilde{o}_2(\v k) \tilde{e}_1(\v k)+\tilde{o}_3(\v k) \tilde{e}_2(\v k) + \tilde{o}_1(\v k) \tilde{e}_4(\v k)= 0 \\
\end{cases}
\end{align*}
We examine the above equation in the neighborhood of $\v k_0$ by expanding $\v k  = \v k_0 + \v q$. 
On the curve $\mathcal{C}_1$ defined in \ref{curve} with $d=3$, for any $r=|\v q|\neq 0$,
\begin{align*}
&\begin{cases}
\tilde{o}_1(\v k) \tilde{e}_3(\v k) = 0 \\
\tilde{o}_1(\v k) \tilde{e}_4(\v k)= 0 \\
\end{cases}
\end{align*}
Because $\tilde{o}_1(\v k)>0$ it implies 
$\tilde{e}_3(\v k) =\tilde{e}_4(\v k)= 0$.  By the continuity of $\tilde{e}_{3,4}(\v k)$ we conclude $\tilde{e}_3(\v k) =\tilde{e}_4(\v k)= 0$ at $r=0$, or in other words, $\tilde{e}_3(\v k_0) =\tilde{e}_4(\v k_0)= 0$. We can repeat this argument by looking at 
$\mathcal{C}_2$ and $\mathcal{C}_3$, which will lead to  $\tilde{e}_1(\v k_0)=\tilde{e}_2(\v k_0)=0$. Combining these results, one gets $A(\v k_0)=0$.

\subsubsection{$G^-(T_+,C_-)$, {\rm or equivalently} $G(Q,T_+)$ {\rm after identifying}  $C$ {\rm with} $Q$}

\be
&C=EII, T=ZII,~~n_0=8,\nn
&\begin{array}{c| c} S(\v k) & \tilde{o}_1 (\v k) YXY + \tilde{o}_2 (\v k)YYI + \tilde{o}_3 (\v k) YZY \\
& + \tilde{o}_4 (\v k) YYX + \tilde{o}_5 (\v k)YYZ + \tilde{o}_6 (\v k)YIY  \\ \hline A(\v k) & \tilde{e}_1 (\v k)YXX + \tilde{e}_2 (\v k)YXZ + \tilde{e}_3 (\v k) YXI \\
&+\tilde{e}_4 (\v k) YYY + \tilde{e}_5 (\v k)YZX + \tilde{e}_6 (\v k)YZZ \\
&+ \tilde{e}_7 (\v k)YZI + \tilde{e}_8 (\v k)YIX + \tilde{e}_9 (\v k)YIZ + \tilde{e}_{10} (\v k)YII  \end{array} 
\label{SAlist}\ee
Here $\{ S(\v k) , A(\v k) \}=0 $ implies

\begin{align}
\begin{cases}
\tilde{e}_{10}(\v k)
\begin{bmatrix}
 \tilde{o}_4(\v k)	\\	 \tilde{o}_5(\v k)	\\	 \tilde{o}_6(\v k)
\end{bmatrix}
= \begin{bmatrix}
-\tilde{e}_6(\v k)	&	-\tilde{e}_8(\v k)	&	\tilde{e}_2(\v k)	\\	
\tilde{e}_5(\v k)	&	-\tilde{e}_9(\v k)	&	-\tilde{e}_1(\v k)	\\	
-\tilde{e}_3(\v k)	&	-\tilde{e}_4(\v k)	&	-\tilde{e}_7(\v k)
\end{bmatrix} \cdot
\begin{bmatrix}
 \tilde{o}_1(\v k)	\\	 \tilde{o}_2(\v k)	\\	 \tilde{o}_3(\v k)
\end{bmatrix} \\
\tilde{e}_{10}(\v k)
\begin{bmatrix}
 \tilde{o}_1(\v k)	\\	 \tilde{o}_2(\v k)	\\	 \tilde{o}_3(\v k)
\end{bmatrix}
= \begin{bmatrix}
-\tilde{e}_6(\v k)	&	\tilde{e}_5(\v k)	&	-\tilde{e}_3(\v k)	\\	
-\tilde{e}_8(\v k)	&	-\tilde{e}_9(\v k)	&	-\tilde{e}_4(\v k)	\\	
\tilde{e}_2(\v k)	&	-\tilde{e}_1(\v k)	&	-\tilde{e}_7(\v k)
\end{bmatrix} \cdot
\begin{bmatrix}
 \tilde{o}_4(\v k)	\\	 \tilde{o}_5(\v k)	\\	 \tilde{o}_6(\v k)
\end{bmatrix}
\end{cases}
\label{weq}
\end{align}
It's straightforward to check that the above equations imply

\begin{align*}
& \left[ \tilde{o}_4^2(\v k)	+	 \tilde{o}_5^2(\v k)	+	 \tilde{o}_6^2(\v k) - \tilde{o}_1^2(\v k)	-	 \tilde{o}_2^2(\v k)	-	 \tilde{o}_3^2(\v k)\right]\tilde{e}_{10}(\v k) =0
\end{align*}

The solutions are 
\begin{align*}
& \tilde{e}_{10}(\v k)=0 \text{ or } \left[  \tilde{o}_4^2(\v k)	+	 \tilde{o}_5^2(\v k)	+	 \tilde{o}_6^2(\v k) -\tilde{o}_1^2(\v k)	-	 \tilde{o}_2^2(\v k)	-	 \tilde{o}_3^2(\v k)\right]=0
\end{align*}
In the following we prove that $ \tilde{e}_{10}(\v k)$ must vanish. \\

The spectral symmetrised Hamiltonian $\tilde{h}(\v k)$ satisfies $\tilde{h}^2(\v k)=\left[S(\v k) + A(\v k)\right]^2=w^2(\v k) III$. We may assume $w(\v k)>0$ without loss of generality. In the following we show that  $\tilde{e}_{10}(\v k)$ must take one of the following values $$\{w(\v k),w(\v k)/2,0,-w(\v k)/2,-w(\v k)\}$$ for each $\v k$.
We first observe that according to \Eq{SAlist} all tensor products in $S(\v k)$ and $A(\v k)$ contain $Y$ as the first factor. Therefore we can factor it out and write $\tilde{h}(\v k)=Y\otimes g(\v k)$ where $g(\v k)$ is a  $4\times 4$ Hermitian matrix function. Next, we express $g(\v k)$ in terms of its eigenbasis, i.e., $g(\v k)= U(\v k)\Lambda(\v k)U^{-1}(\v k)$ where $U(\v k)$ is the basis transformation matrix and $\Lambda(\v k)$ is the diagonal matrix containing the eigenvalues. Under this basis $\tilde{h}^2(\v k)=I\otimes U(\v k)\Lambda^2(\v k) U^{-1}(\v k)$.  Since the spectral symmetrization condition requires  $\tilde{h}^2(\v k)=w^2(\v k) III$, it follows that $$U(\v k)\Lambda^2(\v k) U^{-1}(\v k)=w^2(\v k) II.$$  This implies the eigenvalues of  $\Lambda^2(\v k)$ are four-fold degenerate and are equal to $w^2(\v k)$.  Thus the diagonal elements of $\Lambda(\v k)$ are $\pm w(\v k)$. 
According to \Eq{SAlist} $$\tilde{e}_{10}(\v k)={1\over 8}\Tr[(YII)\tilde{h}(\v k)]={1\over 4}\Tr[\Lambda(\v k)].$$ Because the 
the diagonal elements of $\Lambda(\v k)$ are $\pm w(\v k)$,  $\tilde{e}_{10}(\v k)$ must be equal to one of the five possible values \be \{w(\v k),w(\v k)/2,0,-w(\v k)/2,-w(\v k)\}\label{poss}\ee for each $\v k$.\\

Moreover, because   $\tilde{e}_{10}(\v k)$ is a analytic function of $\v k$ and $w(\v k)>0$ away from $\v k=0$, $\tilde{e}_{10}(\v k)$ can not ``switch track'', i.e., it must be equal to one of above five possible functions throughout the Brillouin zone, away from $\v k=0$. \\

Since $\tilde{h}(\v k)\rightarrow \sum_{j=1}^d k_j \G_j$ as $\v k\rightarrow  0$, it follows that $w(\v k)\rightarrow  |\v k|$ as $\v k\ra 0$. On the other hand, since $\tilde{e}_{10}(\v k)$ is an even function of $\v k$, it must vanishes as an even power in $\v k$ as $\v k\rightarrow 0$, hence
\be
|\tilde{e}_{10}(\v k)|<<w(\v k)~~{\rm as~~}\v k\ra 0.\label{ieq}\ee The only choice in \Eq{poss} that is consistent with \Eq{ieq} is \be
\tilde{e}_{10}(\v k)=0.\ee
\\

Now we may set $\tilde{e}_{10}(\v k)=0$ in  the first three equations of \Eq{weq} and examine these equations in the neighborhood of $\v k_0$ by expanding $\v k  = \v k_0 + \v q$. 
On the curve $\mathcal{C}_1$ defined in \ref{curve} with $d=3$, for any $r=|\v q|\neq 0$,
\begin{align*}
&\begin{cases}
-\tilde{o}_1(\v k) \tilde{e}_6(\v k) = 0 \\
\tilde{o}_1(\v k) \tilde{e}_5(\v k)= 0 \\
-\tilde{o}_1(\v k) \tilde{e}_3(\v k)= 0 \\
\end{cases}
\end{align*}
Because $\tilde{o}_1(\v k)>0$ it implies 
$\tilde{e}_3(\v k) =\tilde{e}_5(\v k)=  \tilde{e}_6(\v k) = 0$.  By the continuity of $\tilde{e}_{3,5,6}(\v k)$ we conclude $\tilde{e}_3(\v k) =\tilde{e}_5(\v k)= \tilde{e}_6(\v k) = 0$ at $r=0$, or in other words, $\tilde{e}_3(\v k_0) =\tilde{e}_5(\v k_0)= \tilde{e}_6(\v k_0) = 0$. We can repeat this argument by looking at $\mathcal{C}_2$ and $\mathcal{C}_3$, 
 which will lead to   $\tilde{e}_4(\v k_0)=\tilde{e}_8(\v k_0)=\tilde{e}_9(\v k_0)=0$ and  $\tilde{e}_1(\v k_0)=\tilde{e}_2(\v k_0)=\tilde{e}_7(\v k_0)=0$. Combining these results, one gets $A(\v k_0)=0$.\\

\subsubsection{$G^-(T_-,C_-)$, {\rm or equivalently} $G(Q,T_-)$ {\rm after identifying}  $C$ {\rm with} $Q$}

\begin{align*}
C=EI, T=ZE,~~n_0=4,~~
\begin{array}{c| c} S(\v k) & \tilde{o}_1 (\v k)YY + \tilde{o}_2 (\v k)IX + \tilde{o}_3 (\v k)IZ  \\ \hline A(\v k) & \tilde{e}_1 (\v k)YI  \end{array} 
\end{align*}
Here $\{ S(\v k) , A(\v k) \}=0 $ implies

\begin{align*}
\begin{cases}
\tilde{o}_1(\v k) \tilde{e}_1(\v k) = 0 \\
\tilde{o}_2(\v k) \tilde{e}_1(\v k) = 0 \\
\tilde{o}_3(\v k) \tilde{e}_1(\v k) = 0 \\
\end{cases}
\end{align*}

We examine the above equations in the neighborhood of $\v k_0$ by expanding $\v k  = \v k_0 + \v q$. 
On the curve $\mathcal{C}_1$ defined in \ref{curve} with $d=3$, for any $r=|\v q|\neq 0$,
\begin{align*}
\tilde{o}_1(\v k) \tilde{e}_1(\v k) = 0 
\end{align*}
Because $\tilde{o}_1(\v k)>0$ it implies 
$\tilde{e}_1(\v k) = 0$.  By the continuity of $\tilde{e}_{1}(\v k)$ we conclude $\tilde{e}_1(\v k) = 0$ at $r=0$, or in other words, $\tilde{e}_1(\v k_0) = 0$. This implies $A(\v k_0)=0$.

\subsubsection{$G(Q,C_-)$}

\be
&&Q=EII, C=ZEI,~~n_0=8,\nn&&
\begin{array}{c| c} S(\v k) & \tilde{o}_1 (\v k)YXY +\tilde{o}_2 (\v k)YZY + \tilde{o}_3 (\v k) IYY \\
	& + \tilde{o}_4 (\v k)IIX + \tilde{o}_5 (\v k)IIZ  \\ 
	\hline 
	A(\v k) & \tilde{e}_1 (\v k)YXX + \tilde{e}_2 (\v k) YXZ + \tilde{e}_3 (\v k)YXI \\
	& + \tilde{e}_4 (\v k)YZX + \tilde{e}_5(\v k) YZZ + \tilde{e}_6(\v k) YZI \\
	&+ \tilde{e}_7 (\v k)IYX + \tilde{e}_8 (\v k) IYZ + \tilde{e}_9 (\v k) IYI + \tilde{e}_{10}(\v k) IIY \end{array} 
\ee

Here $\{ S(\v k) , A(\v k) \}=0 $ implies

\be
	\begin{cases}
		\tilde{o}_1(\v k) \tilde{e}_3(\v k)  +  \tilde{o}_2(\v k) \tilde{e}_6(\v k) + \tilde{o}_3(\v k) \tilde{e}_9(\v k) = 0 \\
		\begin{bmatrix}
		0					&	-\tilde{o}_3(\v k)	& 	\tilde{o}_2(\v k)	\\
		\tilde{o}_3(\v k)	&	0					&	-\tilde{o}_1(\v k)	\\
		-\tilde{o}_2(\v k)	&	\tilde{o}_1(\v k)	&	0					
		\end{bmatrix} 
		\begin{bmatrix}
		\tilde{e}_1(\v k)	\\	\tilde{e}_4(\v k)	\\	\tilde{e}_7(\v k)
		\end{bmatrix}
		= \tilde{o}_5(\v k)\begin{bmatrix}
		\tilde{e}_3(\v k)	\\	\tilde{e}_6(\v k)	\\	\tilde{e}_9(\v k)
		\end{bmatrix}
		\\
		\begin{bmatrix}
		0					&	-\tilde{o}_3(\v k)	& 	\tilde{o}_2(\v k)	\\
		\tilde{o}_3(\v k)	&	0					&	-\tilde{o}_1(\v k)	\\
		-\tilde{o}_2(\v k)	&	\tilde{o}_1(\v k)	&	0					
		\end{bmatrix} 
		\begin{bmatrix}
		\tilde{e}_2(\v k)	\\	\tilde{e}_5(\v k)	\\	\tilde{e}_8(\v k)
		\end{bmatrix}
		= -\tilde{o}_4(\v k)\begin{bmatrix}
		\tilde{e}_3(\v k)	\\	\tilde{e}_6(\v k)	\\	\tilde{e}_9(\v k)
		\end{bmatrix}
		\\
		\begin{bmatrix}
		\tilde{o}_1(\v k)	\\	\tilde{o}_2(\v k)	\\	\tilde{o}_3(\v k)
		\end{bmatrix} \tilde{e}_{10}(\v k)
		= \tilde{o}_4(\v k)
		\begin{bmatrix}
		\tilde{e}_1(\v k)	\\	\tilde{e}_4(\v k)	\\	\tilde{e}_7(\v k)
		\end{bmatrix}
		+ \tilde{o}_5(\v k)
		\begin{bmatrix}
		\tilde{e}_2(\v k)	\\	\tilde{e}_5(\v k)	\\	\tilde{e}_8(\v k)
		\end{bmatrix}
		\\
	\end{cases}
	\label{mxx}
\ee
We examine the above equations in the neighborhood of $\v k_0$ by expanding $\v k  = \v k_0 + \v q$.  
On the curve $\mathcal{C}_1$ defined in \ref{curve} with $d=3$, for any $r=|\v q|\neq 0$, the first equation gives

\begin{align*}
\tilde{o}_1(\v k) \tilde{e}_3(\v k) = 0 \\
\end{align*}
which implies $\tilde{e}_3(\v k) = 0$. By the continuity of $\tilde{e}_3(\v k)$ we conclude $\tilde{e}_3(\v k_0)=0$. We can repeat this argument by looking at 
$\mathcal{C}_2$ and $\mathcal{C}_3$, which  lead to  $\tilde{e}_6(\v k_0)=\tilde{e}_9(\v k_0)=0$. \\

By theorem 1 of \ref{odd_func}, for any radius $|\v q|=r$, we can find a non-self-intersecting closed loop $\gamma_5$, such that (i) $\tilde{o}_5(\v k)=0$ for $\v k\in \gamma_5$, (ii) $\gamma_5$ splits the sphere $|\v q|=r$ into two equal-area regions, and (iii) the antipodal point of any $\v k \in \gamma_5$ is also on $\gamma_5$. Such $\gamma_5$ loops for different radius $r$ form a surface $S_5$ which can be arbitrarily close to $r=0$ (i.e. $\v k_0$). On $S_5$ the second to the fourth lines of \Eq{mxx} gives

\begin{align*}
\begin{bmatrix}
0					&	-\tilde{o}_3(\v k)	& 	\tilde{o}_2(\v k)	\\
\tilde{o}_3(\v k)	&	0					&	-\tilde{o}_1(\v k)	\\
-\tilde{o}_2(\v k)	&	\tilde{o}_1(\v k)	&	0					
\end{bmatrix} 
\begin{bmatrix}
\tilde{e}_1(\v k)	\\	\tilde{e}_4(\v k)	\\	\tilde{e}_7(\v k)
\end{bmatrix}
= 0
\end{align*}

Note that the $3\times 3$ matrix on the left hand side is rank $2$ as long as $\tilde{o}_1(\v k)^2 + \tilde{o}_2(\v k)^2 + \tilde{o}_3(\v k)^2 \neq 0$, which is true for  in the neighborhood of $\v k_0$. This gives the general solution

\begin{align}
\begin{bmatrix}
\tilde{e}_1(\v k)	\\	\tilde{e}_4(\v k)	\\	\tilde{e}_7(\v k)
\end{bmatrix}
= a(\v k)
\begin{bmatrix}
\tilde{o}_1(\v k)	\\	\tilde{o}_2(\v k)	\\	\tilde{o}_3(\v k)
\end{bmatrix}\label{eao}
\end{align}

Note that as $\v k\ra\v k_0$ we can have the following two possibilities: (i) $a(\v k)$ is non-singular, in which case $(\tilde{e}_1(\v k),	\tilde{e}_4(\v k), \tilde{e}_7(\v k))\ra 0$ as $\v k\ra\v k_0$, or (ii) $a(\v k)$ diverges and it compensates for the vanishing magnitude of $(\tilde{o}_1(\v k),	\tilde{o}_2(\v k),	\tilde{o}_3(\v k))$.\\

We first consider possibility (ii).  In this case as $\v k\ra \v k_0$,  $(\tilde{e}_1(\v k),	\tilde{e}_4(\v k), \tilde{e}_7(\v k))$ can be non-zero. However, its direction must be parallel (or antiparallel) to $$\hat{n}(\v k)=(\tilde{o}_1(\v k),	\tilde{o}_2(\v k),	\tilde{o}_3(\v k))/|(\tilde{o}_1(\v k),	\tilde{o}_2(\v k),	\tilde{o}_3(\v k))|.$$ 
Let's look at the pair of antipodal points on a $\gamma_5$ loop at an infinitesimal radius $|\v q|=r$. 
By continuity of $(\tilde{o}_1(\v k),\tilde{o}_2(\v k),\tilde{o}_3(\v k))$ and $\hat{n}(\v k)$ must change continuously on $\gamma_5$. This implies $\hat{n}(\v k)\cdot(\tilde{e}_1(\v k),\tilde{e}_4(\v k),\tilde{e}_7(\v k))$ changes continuously on $\gamma_5$.
Since $\hat{n}(\v k)$ is odd and $\tilde{e}_{1,4,7}(\v k)$ are even, $\hat{n}(\v k)\cdot(\tilde{e}_1(\v k),\tilde{e}_4(\v k),\tilde{e}_7(\v k))$ has opposite sign among antipodal points on $\gamma_5$. Thus it must vanish at some intermediate point $\v k'$ on $\gamma_5$. Since $(\tilde{e}_1(\v k),\tilde{e}_4(\v k),\tilde{e}_7(\v k)) \parallel \hat{n}(\v k)$ on $\gamma_5$ by \eqref{eao}, thus $(\tilde{e}_1(\v k'),\tilde{e}_4(\v k'),\tilde{e}_7(\v k'))=\v 0$. 
By connecting such point for different $r$, we arrive at a continuous path on which $(\tilde{e}_1(\v k),\tilde{e}_4(\v k),\tilde{e}_7(\v k))=\v 0$. By continuity we have 
\begin{align*}
(\tilde{e}_1(\v k_0),\tilde{e}_4(\v k_0),\tilde{e}_7(\v k_0))=0
\end{align*}

%
We can repeat the same arguments for the surface corresponds to  $\tilde{o}_4(\v k)=0$. This lead to $(\tilde{e}_2(\v k_0),	\tilde{e}_5(\v k_0), \tilde{e}_8(\v k_0))=\v 0$.\\

Moreover, by the theorem 2 of \ref{odd_func}, on the sphere correspond to any  $r=|\v q|$, one can find a point $\v k$ such that both $\tilde{o}_4(\v k)$ and $\tilde{o}_5(\v k)$ are zero. Such points for different $r$ form a curve which approaches $\v k_0$ as $r\ra 0$. On the curve, the last of \Eq{mxx} gives

\begin{align*}
\begin{bmatrix}
\tilde{o}_1(\v k)	\\	\tilde{o}_2(\v k)	\\	\tilde{o}_3(\v k)
\end{bmatrix} \tilde{e}_{10}(\v k) 
= 0
\end{align*}

Since on this curve since $\tilde{o}_1(\v k),	\tilde{o}_2(\v k),	\tilde{o}_3(\v k)$ cannot simultaneously be zero, it follows that $\tilde{e}_{10}(\v k) =0$ on the curve. Due to the continuity of  $\tilde{e}_{10}(\v k)$ we conclude that  $\tilde{e}_{10}(\v k_0) =0$. Combining all of the above results, we conclude $A(\v k_0)=0$.

\subsubsection{$G^+(Q, T_- ,C_-)$}

\be
&&Q=EII, T=ZEI, C=ZIE,~~n_0=8,\nn&&
\begin{array}{c| c} S(\v k) & \tilde{o}_1 (\v k)YYX + \tilde{o}_2 (\v k)YYZ + \tilde{o}_3 (\v k)IXI + \tilde{o}_4 (\v k)IZI  \\ \hline A(\v k) & \tilde{e}_1 (\v k) YIX + \tilde{e}_2 (\v k)YIZ +\tilde{e}_3 (\v k)IXY + \tilde{e}_4 (\v k)IZY \end{array} 
\ee
Here $\{ S(\v k) , A(\v k) \}=0 $ implies

\be
\begin{cases}
\tilde{o}_2(\v k) \tilde{e}_4(\v k)  +  \tilde{o}_3(\v k) \tilde{e}_1(\v k)  = 0 \\
\tilde{o}_1(\v k) \tilde{e}_4(\v k)  -  \tilde{o}_3(\v k) \tilde{e}_2(\v k)  = 0 \\
\tilde{o}_1(\v k) \tilde{e}_1(\v k)  +  \tilde{o}_2(\v k) \tilde{e}_2(\v k)  = 0 \\
\tilde{o}_2(\v k) \tilde{e}_3(\v k)  -  \tilde{o}_4(\v k) \tilde{e}_1(\v k)  = 0 \\
\tilde{o}_1(\v k) \tilde{e}_3(\v k)  +  \tilde{o}_4(\v k) \tilde{e}_2(\v k)  = 0 \\
\tilde{o}_3(\v k) \tilde{e}_3(\v k)  +  \tilde{o}_4(\v k) \tilde{e}_4(\v k)  = 0 \\
\end{cases}
\label{mxxx}
\ee
We examine the above equations in the neighborhood of $\v k_0$ by expanding $\v k  = \v k_0 + \v q$. 
On the curve $\mathcal{C}_1$ defined in \ref{curve} with $d=3$, for any $r=|\v q|\neq 0$, the second and third lines of \Eq{mxxx} give 

\begin{align*}
&\begin{cases}
\tilde{o}_1(\v k) \tilde{e}_4(\v k)    = 0 \\
\tilde{o}_1(\v k) \tilde{e}_1(\v k)    = 0 \\
\end{cases}
\end{align*}
which implies $\tilde{e}_1(\v k) =\tilde{e}_4(\v k)= 0$. By the continuity of $\tilde{e}_{1,4}(\v k)$ we conclude $\tilde{e}_1(\v k_0)=\tilde{e}_4(\v k_0)=0$. We can repeat this argument by looking at 
$\mathcal{C}_2$ and $\mathcal{C}_3$, which will lead to $\tilde{e}_2(\v k_0)=0$. \\

It remains to prove that $\tilde{e}_3(\v k_0) = 0$. By the theorem 1 in \ref{odd_func}, for any radius $r=|\v k|$ one can find a non-self-intersecting closed loop $\gamma_4$ such that $\tilde{o}_4(\v k_0) = 0$. As a function of $r$ all such loops span surface which approach $\v k_0$ as $r\ra 0$. Everywhere on the surface, the 4-6 lines of \Eq{mxxx} give

\begin{align*}
\begin{cases}
\tilde{o}_2(\v k) \tilde{e}_3(\v k)  = 0 \\
\tilde{o}_1(\v k) \tilde{e}_3(\v k)  = 0 \\
\tilde{o}_3(\v k) \tilde{e}_3(\v k)  = 0 \\
\end{cases}
\end{align*}

Because $(\tilde{o}_1(\v k) , \tilde{o}_2(\v k),\tilde{o}_3(\v k))$ has non-trivial mapping degree around $k_0$, they cannot be simultaneously zero. It follows that $\tilde{e}_3(\v k)  = 0$ everywhere on the surface. By the continuity of  $\tilde{e}_3(\v k)$, we conclude that  $\tilde{e}_3(\v k_0)  = 0$. Combining these results, one gets $A(\v k_0)=0$.

\section{Odd continuous functions on $S^2$}\label{odd_func}

In this appendix, we prove some properties for odd continuous functions obeying  $o(-\v q) = - o(\v q)$, on a two-sphere $S^2$ formed by $|\v q| =$ constant.

\subsection{Theorem 1}

\begin{blockquote}\
	{\bf Theorem 1} For any continuous odd function $o(\v q)$ defined on a sphere formed by $|\v q|=r$, there exists a non-self intersecting closed loop $\gamma_o$ on the sphere, such that (i) $o(\v q)=0$ for $\v q\in \gamma_o$, (ii) the curve separates the sphere into two equal-area regions, and (iii) the antipodal point of any point $\v q$ on the loop also belongs to the loop.
\end{blockquote}

\underline{Proof}:
We will prove it by explicitly constructing  $\gamma_o$. If $o(\v q)=0$ everywhere on the sphere, any arbitrary great circle on $S^2$ can be used for $\gamma_o$. Thus the non-trivial case must have at least one point, $\v q_*$, such that $o(\v q_*)\neq0$. Without loss of generality, let's assume $o(\v q_*)>0$. Due to the oddness, $o(-\v q_*)<0$. Now consider a geodesic (or a great arc) connecting $\v q_*$ and $-\v q_*$.  Owing to the continuity of $o(\v q)$, the function must change sign an odd number of times as the geodesic is traversed. The points at which the sign changes take place must correspond to  $o(\v q)=0$. They can either be discrete points or form a continuous segment on the great arc. In either case we can choose a middle point (which can either be the mid point of the middle zero-segment, or just the mid point among the discrete points where the sign change takes place). We then rotate the great arc through the whole $2 \pi$ angle. As a function of angle, the aforementioned mid points span the loop $\gamma_o$. The loop can not self-intersect because we only choose a single point on every great arc.\\

Moreover, 
due to the oddness of $o(\v q)$, the mid point $\v q_m$ chosen for a given great arc must be antipodal to $-\v q_m$ chosen on the complementary great arc (a great arc and its complementary form a great circle). This guarantees that the loop $\gamma_o$ will separate the sphere into two regions with equal areas. By construction, the antipodal point of any point $\v q$ on the $\gamma_o$ is also on the loop.  Q.E.D..

\subsection{Theorem 2}

\begin{blockquote}\
	{\bf Theorem 2} For any two continuous odd functions $o_1(\v q), o_2(\v q)$ defined on a sphere $|\v q| =r$, there exists at least a point $\v q_{**}$ such that $o_1(\v q_{**})=o_2(\v q_{**})=0$.
\end{blockquote}

\underline{Proof}:
Assuming the opposite, namely, there is no point $\v q$ at which $o_1(\v q)=o_2(\v q)=0$. For these two functions $o_1(\v q), o_2(\v q)$, we can use  theorem 1 to find the non-self-intersecting closed loops $\gamma_{1}$ and $\gamma_{2}$ which separately divide the sphere into two equal-area regions, and $o_1(\v q)=0$ for $\v q\in\gamma_1$ and $o_2(\v q) = 0$ for $\v q\in\gamma_2$. $\gamma_1$ and $\gamma_2$ must not intersect each other, otherwise the intersection will satisfy $o_1(\v q)=o_2(\v q)=0$. Thus, one loop must be totally enclosed by the other loop, which contradicts the statement that they separately split the sphere into two equal-area regions. Q.E.D.

\newpage

{\noindent\bf Acknowledgement} \\

\noindent This work was primarily supported by the Theory Program at the Lawrence Berkeley National Laboratory, which is funded by the U.S. Department of Energy, Office of Science, Basic Energy Sciences, Materials Sciences, and Engineering Division under Contract No. DE-AC02-05CH11231. This research is also funded in part by the Gordon and Betty Moore Foundation's EPIQS Initiative, Grant GBMF8688 to DHL. \\

{\noindent\bf References} 

\bibliographystyle{ieeetr}
\bibliography{bibs}

\end{document}